\documentclass[aps,prx,twocolumn,showpacs,superscriptaddress,floatfix]{revtex4-1}
\usepackage{dcolumn}          
\usepackage[dvips]{graphicx}  
\usepackage[english]{babel}
\usepackage{color,amsmath,amssymb,amsbsy,amscd,bm}
\usepackage{epsf,epsfig}
\usepackage{tikz}	
\usetikzlibrary{arrows,shapes,snakes}
\usetikzlibrary{backgrounds,fit,decorations.pathreplacing}  

\newcommand{\ket}[1]{\ensuremath{\left| #1\right\rangle}} 
\newcommand{\bra}[1]{\ensuremath{\left\langle #1\right|}} 

\DeclareGraphicsExtensions{.eps}

\begin{document}
\preprint{}
\title[]{Delocalized Quantum States Enhance Photocell Efficiency}
\author{Yiteng Zhang}
\affiliation{Qatar Environment and Energy Research Institute, Qatar Foundation, Doha, Qatar}
\affiliation{Department of Chemistry, Physics and Birck Nanotechnology Center, 
             Purdue University, West Lafayette, IN 47907 USA}
\author{Sangchul Oh}
\affiliation{Qatar Environment and Energy Research Institute, Qatar Foundation, Doha, Qatar}
\author{Fahhad H. Alharbi}
\affiliation{Qatar Environment and Energy Research Institute, Qatar Foundation, Doha, Qatar}
\author{Greg Engel}
\affiliation{Department of Chemistry, University of Chicago, Chicago, IL, 60637 USA}
\author{Sabre Kais}
\email{kais@purdue.edu}
\affiliation{Qatar Environment and Energy Research Institute, Qatar Foundation, Doha, Qatar}
\affiliation{Department of Chemistry, Physics and Birck Nanotechnology Center, 
             Purdue University, West Lafayette, IN 47907 USA}
\date{\today}
\begin{abstract}
The high quantum efficiency of photosynthetic complexes has inspired researchers to explore new 
routes to utilize this process for photovoltaic devices. Quantum coherence has been demonstrated 
to play a crucial role within this process. Herein, we propose a three-dipole system as a model 
of a new photocell type which exploits the coherence among its three dipoles. We have proved 
that the efficiency of such a photocell is greatly enhanced by quantum coherence. We have also 
predicted that the photocurrents can be enhanced by about 49.5\% in such a coherent coupled  dipole
system compared with the uncoupled dipoles. These results suggest a promising novel design aspect 
of photosynthesis-mimicking photovoltaic devices.
\end{abstract}
\maketitle
\section{Introduction}
Long-lived quantum coherence been observed in photosynthesis after laser excitation~\cite{Engel2007,
Calhoun2009, Abramavicius2009,Panitchayangkoon2010,Harel2010,Hayes2013,Romero2014}. It has attracted 
much attention on how quantum coherence could be enhanced in complex biological environment and 
how it may play a key role in efficient exciton transports~\cite{Mohseni2008,Plenio2008,Rebentrost2009,
Zhu2011,Yeh2012}. It is well known that the photon-to-charge conversion quantum efficiency of 
photosynthesis in plants, bacteria, and algae can be almost 100\% under certain conditions. 
While photosynthesis converts sunlight into chemical energy, solar cell converts sunlight into 
electric energy. According to Shockley and Queisser, the efficiency of photovoltaic energy conversion 
is limited to 33\%, based on the energy band gap and solar spectrum, due to the radiative 
recombination of electron-hole pairs, thermalization, and unabsorbed photons~\cite{Shockley1961}. 
Various attempts have been made to improve the performance of photovoltaic 
devices~\cite{Wurfel2009,Wurfel2011,Miller2012,Alharbi2013,Alharbi2014}. Mimicking photosynthesis 
presents a promising route by which to increase the efficiency of the current 
solar cell technology~\cite{Blankenship2011}. Consequently, there has been a long-standing and 
increasing interest in the understanding of the physics describing the energy conversion within 
photosynthesis.  Recently, quantum coherence has demonstrated its crucial role 
in the energy conversion during photosynthesis~\cite{Engel2007,Calhoun2009,Abramavicius2009,
Panitchayangkoon2010,Harel2010,Hayes2013,Romero2014,Mohseni2008,Plenio2008,Rebentrost2009,
Zhu2011,Yeh2012}. Similarly, it has been shown that quantum coherence can be used to alter 
the conditions of the detailed balance and thereby enhance the quantum efficiency 
in photocell~\cite{Scully2010,Scully2011,Svidzinsky2011,Dorfman2013}. In principle, 
the Shockley-Queisser model is a two-extended-level model. By incorporating more levels and 
tuning them carefully, the conversion efficiency can be improved.

Recently, Creatore {\it et al.}~\cite{chin} have shown that the delocalized quantum state is capable of 
improving the photocurrent of a photocell by at least 35\% in compared with a photocell with 
the localized quantum state when treating the photon-to-charge 
conversion as a continuous Carnot-like cycle~\cite{note}. Within their model, the two delocalized states, 
called the bright and dark states, of the dipole-dipole interacting two donors play a key role 
in improving the efficiency of the PV cell. Due to the constructive interference, the optical 
transition rate between the ground and the bright states becomes two times stronger than 
the uncoupled donor case. While it is blocked through the bright state due to the destructive 
interference, the electron transition from the excited donor to the acceptor is made only through 
the dark state and its rate is two times larger than the uncoupled donor case, due to the 
constructive interference. Consequently, the presence of quantum coherence of the delocalized donor 
states alters the conditions for the thermodynamic detailed balance; that results in 
the enhancement of the efficiency of the photocell.

In this paper, we show that a photocell with three suitably arranged electron donors coupled via 
dipole-dipole interactions can result in an enhancement of photocurrents 
by about 49.5\% over a classical photocell. While inspired by Creatore {\it et al.}~\cite{chin},
our three coupled donors, rather than the two coupled ones, makes another big improvement in the 
efficiency of a PV cell. The origin of the photocurrent enhancement is explained by the key roles 
of the delocalized excited states of the donor system. The dipole-dipole coupling between donors 
make three degenerate and localized one-exciton levels split into three delocalized levels:
the bright, almost-dark, and dark states. The photon absorption and emission rates between the 
ground and the bright excited state becomes about 2.91 times larger than that of the uncoupled 
donor case, which is due to the constructive interference of three donors. 
While the electron transferring from the donor to the acceptor through the almost-dark state is 
enhanced by about 2.91 times compared to the uncoupled donor case, but is almost blocked 
through the bright state, which are also due to the constructive and destructive interferences
of the delocalized donor states. Basically, essential physics of our triple-donor model is
similar to that of Creatore {\it et al.}'s two donor model, but more enhanced by 
collective properties. While it seems challenging, our proposed model could be realized 
by nanotechnologies inspired by natural light-harvesting structures.


\section{PV Models with Two Donors}
Before introducing a photovoltaic cell model with three donors, we discuss in detail 
how a configuration of two dipoles moments of two donor affects the efficiency of a PV cell
in Creatore {\it et al.}'s model~\cite{chin}. The excitation of a molecule is simply modeled 
as a two-level system with the ground state $\ket{b}$ and excited state $\ket{a}$. 
The optical transition between them is characterized by the optical dipole moment 
$\pmb{\mu} = e\bra{a}\pmb{r}\ket{b}$. For a molecular aggregate composed of electric neutral 
molecules, the intermolecular interaction is given by the electrostatic dipole-dipole 
coupling~\cite{Abramavicius2009}
\begin{equation}
J_{12} = \frac{1}{4\pi\epsilon\epsilon_0}
      \left( \frac{\pmb{\mu}_1\cdot\pmb{\mu}_2}{r^3}
            - \frac{3(\pmb{\mu}_1\cdot\pmb{r})(\pmb{\mu}_2\cdot\pmb{r})}{r^5} \right),
\label{dipole-dipole}
\end{equation}
where dipole moment $\pmb{\mu}_1$ is located at $\pmb{r}_1$, $\pmb{\mu}_2$ at $\pmb{r}_2$,
and $\pmb{r}= \pmb{r}_2 - \pmb{r}_1$ is the radius vector from $\pmb{\mu}_1$ to $\pmb{\mu}_2$. 
Typically, the strength of $J_{12}$ is much weaker than the excitation energy $\hbar\omega 
= E_{a} -E_b$. The exciton dynamics of the aggregate is described by Hamiltonian~\cite{note1}
\begin{equation}
H = \sum_i \hbar\omega_{i}\sigma_i^{+}\sigma_i^{-} 
 + \sum_{i\ne j} J_{ij}(\sigma_i^{-}\sigma_{j}^{+} + h.c.)
\label{Hamil_exciton}
\end{equation}
where $\sigma^+ = \ket{a}\bra{b}$ and $\sigma^- = \ket{b}\bra{a}$ are the Pauli raising and 
lowering operators, respectively.

According to Eq.~(\ref{dipole-dipole}) the strength of $J_{12}$ depends on how dipole moments 
are aligned. In Creatore {\it et al.}'s paper~\cite{chin}, the donor is a dimer where the dipole 
moment $\pmb{\mu}_1$ is always perpendicular to the radius vector $\pmb{r}$ so the second term 
in Eq.~(\ref{dipole-dipole}) vanishes. The dipole-dipole coupling is given by $J_{12} 
\propto \pmb{\mu}_1\cdot\pmb{\mu}_2 = \mu_1\mu_2\cos\theta$ with angle $\theta$ between 
two dipole moments. This gives rise to the simple angle-dependence energy gap 
$\Delta E = 2J^{0}_{12}|\cos\theta|$ between the symmetric and antisymmetric excited states. 
The spontaneous decay rates are also proportional to $|\mu|^2(1\pm\cos\theta)$.

\begin{figure}
\includegraphics[width=0.45\textwidth]{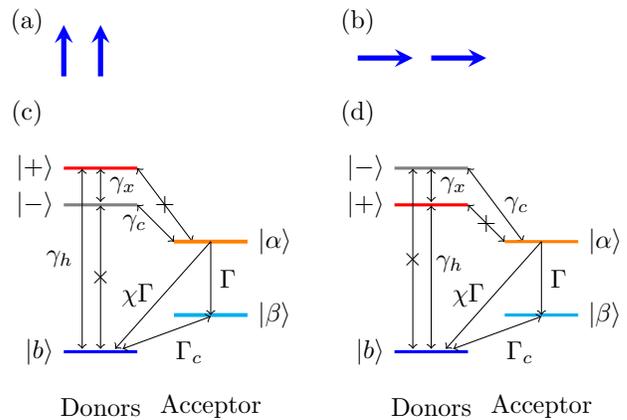}
\caption{Alignments of two dipole moments for (a) H-aggregate and (b) J-aggregate.
Energy level diagrams and electron transition paths of (c) H-aggregate and (d) J-aggregate.
The symmetric state $\ket{+}$ is optically bright and has the absorption and emission rate 
$\gamma_h$, but has no electron transition channel to the donor. The antisymmetric state 
$\ket{-}$ is dark but has the electron transfer path to the donor. The electron transition rate 
$\gamma_x$ between the bright state $\ket{+}$ and dark state $\ket{-}$ is caused by 
thermal phonons.
}
\end{figure}

\begin{figure}[htpb]
\begin{center}
\includegraphics[width=0.5\textwidth]{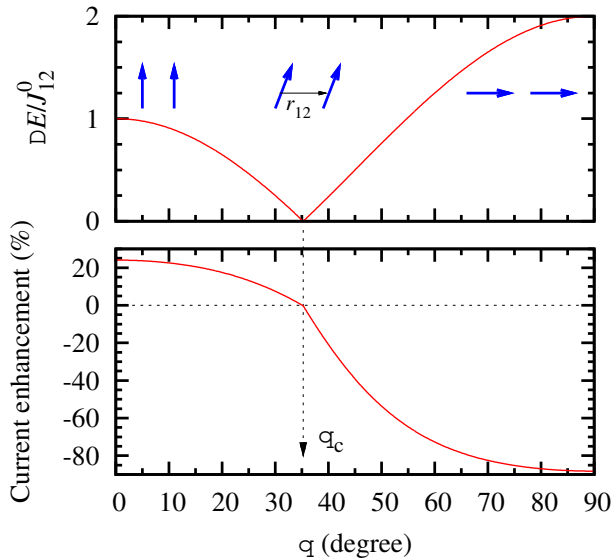}
\end{center}
\caption{(a) The energy gap $\Delta E$ between the bright and dark states and (b)
the current enhancement as a function of the tilt angle $\theta$ with respect to 
the vertical axis. In (a) the alignment of two dipole moments is shown as the blue arrows.
The two parallel dipole moments are aligned in head to head manner (H-aggregate) at
$\theta=0$ and in head to tail manner (J-aggregate) at $\theta=\pi/2$.
In (b) the black arrow points to the magic angle $\theta_c\approx 35.26^{\circ}$.}
\label{mangle}
\end{figure}

Molecules in aggregates, however, are more probable to be aligned collectively, not 
independently. We study how the H and J aggregate donor alignments affect the efficiency 
of PV cells. As illustrated in Fig. 1, we consider the two dipole moments tilted at the same 
angle $\theta$ with respect to the vertical axis. The angle dependence of the dipole-dipole 
coupling of Eq.~(\ref{dipole-dipole}) becomes
\begin{equation}
J_{12}(\theta) = J_{12}^{0}\left( 1 -3\cos^2(\tfrac{\pi}{2} - \theta)\right).
\end{equation}
This implies the angle dependence of the Davidov energy splitting $\Delta E(\theta) = 2|J_{12}(\theta)|$
between the symmetric and antisymmetric states and explain the transition between the H-aggregate 
and the J-aggregate at the magic angle $\theta_c=\cos^{-1}(\tfrac{1}{\sqrt{3}})\approx 54.74^{\circ}$
when the angle is measured from $\pmb{r}$. Here the angle is measured with respect to the vertical axis
so one has the magic angle $\theta_c \approx  35.26^{\circ}$ as shown in Fig.~\ref{mangle}.

In contrast to Creatore {\it et al.}'s configuration, the symmetric state 
$\ket{+}=\frac{1}{\sqrt{2}}(\ket{a_1} + \ket{a_2})$ in our model is always an optically 
active state (bright state)~\cite{note2}. For angles less than $\theta_c$, this level is higher 
than the antisymmetric (dark) state $\ket{-} = \frac{1}{\sqrt{2}}(\ket{a_1} -\ket{a_2})$ so 
the optical transition is shifted to the blue (H-aggregate). On the other hand, for angles 
greater than $\theta_c$, the antisymmetric state is higher so the optical transition is 
changed to the red (J-aggregate). Note that classically the total dipole moment 
is always $2|\mu|$ because the two dipole moments point to the same direction.
The dipole matrix element between the ground and bright states is $\sqrt{2}|\mu|$
so the optical transition rate $\gamma_h$, proportional to the square of the dipole matrix 
element, becomes doubled, i.e, $\gamma_h=2\gamma_{1h}$, in compared with an uncoupled 
donor case. We calculate how the current enhancement is dependent on the angle $\theta$, 
as plotted in Fig.~\ref{mangle}. The two energy levels $E_{\pm}$ for the symmetric and 
antisymmetric states move to $(E_a -E_b)/2$ as the angle $\theta$ increases.
This affects the Bose-Einstein distributions, $n_{h}$ of thermal photons, 
$n_x$ and $n_c$ of thermal phonon through the gaps $E_+ -E_b$, $E_+ -E_-$, 
and $E_\pm -E_{\alpha}$, respectively. Because of $(E_a-E_b) \gg J$, the distribution $n_h$
is dependent little on the angle $\theta$. However, $n_x$ and $n_c$ are 
strongly affected by the angle $\theta$ so drastic changes in current enhancement. 
For H-aggregate case $(\theta < \theta_c)$, $n_x$ increases but $n_c$ decreases 
as angle $\theta$ increases. So the the current enhancement decreases as angle $\theta$ 
increases. For J-aggregate case $(\theta > \theta_c)$, the bright state
is lower than the dark state. An electron in the bright state jumps to the dark state
via only the absorption $\gamma_x$ of thermal phonons (in H-aggregate case, the transition
from the bright to dark states can be done via the stimulated emission and spontaneous emission of thermal 
phonons, $\gamma_x(1+n_x)$). Thus, the transition from the donor to acceptor is very low, and
the current enhancement is negative as shown in Fig.~\ref{mangle}.
 In our model as well as Creatore {\it et al.}'s 
model~\cite{chin}, two donors are coupled to the acceptor so the bright state has no electron 
transferring channel to the acceptor because of the destructive interference. 
If a donor system is composed of many molecules (for example a linear chain), it is likely 
that only some donor molecules (or the molecules at the end site) are coupled to 
the acceptor so the transition path of the bright state to the donor would not 
be blocked.

\section{PV model with three dipole donors}
\subsection{Model}

\begin{figure}[htpb]
\includegraphics[width=0.4\textwidth]{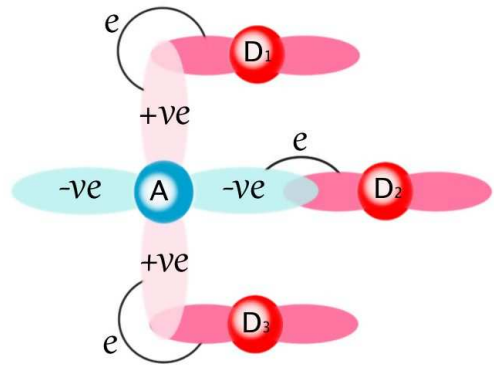}
\caption{Schematics of our PV cell. Three optically active donors, denoted by $D_1, D_2, D_3$,
become excited by absorbing incident photons and their excited electrons are transferred to 
the acceptor $A$. The pink and blue shadowed regions surrounding the molecules represent
the molecular orbitals representing the spatial distribution of electron density.}
\label{model}
\end{figure}

The photocell model, proposed here, is depicted in Fig. \ref{model}. The picture of a classical cyclic 
engine is described as the following: $D_1$, $D_2$, and $D_3$ represent three identical and initially 
uncoupled donor molecules which are aligned around an acceptor molecule $A$. Initially, the system 
starts in the ground state $\ket{b}$. The cycle of electron transport begins with the absorption of solar 
photons populating the uncoupled donor excited states $\ket{a_1}$, $\ket{a_2}$, and $\ket{a_3}$. 
Then the excited electrons can be transferred to the acceptor molecule, the charge-separated state 
$\ket{\alpha}$, with any excess energy radiated as a phonon. The excited electron is then assumed to be 
used to perform work, leaving the charge-separated state $\ket{\alpha}$ decaying to the sub-stable 
state $\ket{\beta}$. The recombination between the acceptor and the donor is also considered with 
a decay rate of $\Gamma_{\alpha\to b}=\chi\Gamma$, where $\chi$ is a dimensionless fraction. This loss 
channel brings the system back into the ground state without producing a work current, which could be 
a significant source of inefficiency. Finally, the state $\ket{\beta}$ decays back to the charge neutral 
ground state, closing the cycle.  If considering the quantum effects resulting from the long-range 
dipole-dipole interaction, the new element of the system is the formation of new optically excitable 
states through strong exciton coupling among the donor molecules~\cite{chin}.

\begin{figure}[htbp]
\centering{\includegraphics[width=0.4\textwidth]{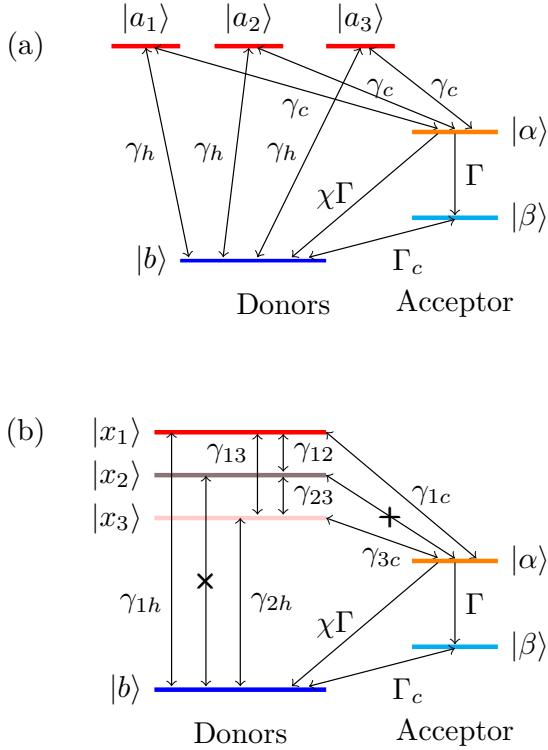}}
\caption{Energy levels of the PV models (a) with the acceptor 
and three uncoupled donors and (b) with the acceptor and three dipole-dipole coupled donors. 
Black arrows indicate possible electron-transition paths. In (a) all the three donors are 
uncoupled and identical so have the same excitation energies ($E_i$), the same the photon 
absorption and emission rates $\gamma_h$ between the ground state $\ket{b}$ 
and excited states $\ket{a_i}$, and the same electron transfer rates $\gamma_c$
between the excited donors ($\ket{a_i}$) and the acceptor ($\ket{\alpha}$).
In (b) due to the dipole-dipole couplings between three donors, the three degenerate excited levels 
in (a) become split, denoted by $\ket{x_i}$. The dark level ($\ket{x_2}$) is optically 
forbidden and has no electron transfer path to the donor ($\ket{\alpha}$). 
}
\label{states2}
\end{figure}

For simplicity, we assume that three donors ($D_1,D_2$ and $D_3$) are identical and degenerate,
so the uncoupled excited states $\ket{a_1}, \ket{a_2}$ and $\ket{a_3}$ of the three donors
have the same excitation levels $E_{1} = E_{2} = E_{3}=\hbar\omega$. 
Furthermore, their dipole moments are aligned to the same direction, 
$\pmb{\mu}_1 = \pmb{\mu}_2 = \pmb{\mu}_3 = \pmb{\mu}$ as depicted in Fig.~\ref{model}.
We assume the dipole-dipole interaction between only nearest neighbors. The dipole-dipole 
couplings between $D_1$ and $D_2$, and $D_2$ and $D_3$ are denoted by $J$,
but there is no coupling between $D_1$ and $D_3$. The Hamiltonian for the system 
of three interacting donors is written as
\begin{equation}
\label{threeH}
H = \sum_{i=1}^{3}\hbar\omega\sigma_i^{+}\sigma_i^{-} 
  + J(\sigma_1^{-}\sigma_2^{+} + \sigma_2^{-}\sigma_3^{+} + h.c.)\,.
\end{equation}
It is straightforward to obtain the three single-excitation states of Hamiltonian~(\ref{threeH}):
$\ket{x_1} = \frac{1}{2}(\ket{a_1} +\sqrt{2}\ket{a_2} + \ket{a_3})$, 
$\ket{x_2} = \frac{1}{\sqrt{2}}(\ket{a_1} -\ket{a_3})$, 
and $\ket{x_3} = \frac{1}{2}(\ket{a_1}-\sqrt{2}\ket{a_2} +\ket{a_3})$.
The corresponding eigenvalues are obtained as 
$E_{x_1} = E+\sqrt{2}J, E_{x_2} = E$, and $E_{x_3} = E-\sqrt{2}J$. 

The dipole moment between the state $\ket{x_1}/\ket{x_3}$ and the ground state $\ket{0}$  
is enhanced/weakened by constructive interference between the individual 
transition dipole matrix elements, $\mu_{x_1/x_3} = \frac{1}{2}(\mu_1\pm\sqrt{2}\mu_2+\mu_3)
=(1\pm\frac{1}{\sqrt{2}})\mu$, while the dipole moment of the state $\ket{x_2}$ cancels due to 
destructive interference. This means the state $\ket{x_2}$, comprised of the antisymmetric 
combination of the uncoupled $\ket{a_1}$ and $\ket{a_3}$ states, describes an optically 
forbidden dark state. On the contrary, the $\ket{x_1}$ and $\ket{x_3}$ states describe 
two optically active bright states with photon absorption and emission rates 
$\gamma_{1h} \propto |\mu_{x_1}|^2=(\tfrac{3}{2}+\sqrt{2})|\mu|^2$ and 
$\gamma_{3h} \propto |\mu_{x_3}|^2=(\tfrac{3}{2}-\sqrt{2})|\mu|^2$, respectively,
in compared with the uncoupled case, $\gamma_h \propto |\mu|^2$.
In other words, $\ket{x_1}$ is much brighter than $\ket{x_3}$, as the photon absorption and emission rate of 
$\ket{x_1}$ is enhanced while that of $\ket{x_3}$ is weakened. 
Obviously, the dark state $\ket{x_2}$ has a resultant charge transfer matrix element equal to zero. 

The intermolecular dipole interaction will also modify the transition rate between the donors 
and acceptor. The electron transfer matrix elements leading to charge separation have been chosen 
to have the same magnitudes $|t_{D_1A}|=|t_{D_2A}|=|t_{D_3A}|=t$. Also, we assume that the acceptor 
molecule hosts an electron within its lowest unoccupied molecular orbital, which is characterized 
by the shape of the $d$-orbitals (See Fig. \ref{model}). We have also assumed that the donor molecules 
are located close to different lobes of the acceptor molecular orbital; this leads to electron 
transfer matrix elements with the same magnitudes but different signs, i.e., 
$t_{D_1A}=-t_{D_2A}=t_{D_3A}=t$. Due to effects of the dipole-dipole interactions, 
the eigenstates of the three optically excited donors are no longer uncoupled, but are coherent 
exciton states.  The bright states $\ket{x_1}/\ket{x_3}$ have matrix elements 
$t_{x_1A/x_3A}=\tfrac{1}{2}( t_{D_1A} \mp\sqrt{2}t_{D_2A} +t_{D_3A})
=(1\mp\frac{1}{\sqrt{2}})t$, giving decreased/enhanced electron transfer rates of 
$\gamma_{1c/3c}\propto|t_{x_1A/x_3A}|^2=(\frac{3}{2}\mp \sqrt{2})|t|^2$,
in compared with the uncoupled case $\gamma_c\propto|t|^2$. 
These modifications of electron transfer matrix elements play a crucial role in the enhancement 
of photocurrents within our photocell model.
The assumptions surrounding the electron transfer matrix elements is identical to that in 
Ref.~\onlinecite{chin}. 

Another crucial procedure in our model is phonon-mediated energy relaxation, which can be very 
effective between exciton states with strong pigment overlap~\cite{chin,excite}. These relaxations 
are included in our kinetic model via the relaxation rates $\gamma_{12}, \gamma_{13}, \gamma_{23}$. 
Assuming that the new donor states are directly populated by the absorption of weak incoherent solar 
photons, the kinetics of the optically excited states obey the Pauli master equation (PME) by treating 
the donor-light, electron transfer, and bright-dark relaxation coupling by second-order 
perturbations~\cite{chin}.  

\begin{widetext}
The PMEs for the uncoupled case, describing the processes as shown in Fig.~\ref{states2} (a), 
are written as
\begin{equation}
\begin{split}
\label{e1}
\dot{p}_{1}
&=\gamma_{h}[n_{h} p_b - (1+n_{h})p_{1}] 
 +\gamma_c[n_c p_{\alpha} -(1+n_c)p_{1}],\\[5pt]
\dot{p}_{2}
&=\gamma_h[n_hp_b -(1+n_h)p_{2}] 
 +\gamma_c[n_c p_{\alpha} - (1+n_c)p_{2}],\\[5pt]
\dot{p}_{3}
&=\gamma_h[n_hp_{b} -(1+n_h)p_{3}] 
 +\gamma_c[n_c p_{\alpha} -(1+n_c)p_{3}],\\[5pt]
\dot{p}_{\alpha}
&=\gamma_c(1+n_c)(p_{1} + p_{2} + p_{3}) 
 -3\gamma_c n_cp_{\alpha}
  - \Gamma(1+\chi)p_{\alpha},\\[5pt]
\dot{p}_{\beta}
&= \Gamma_c[N_c p_{b} - (1+N_c)p_{\beta}] + \Gamma p_{\alpha}
\end{split}
\end{equation}
where we use the notation $p_i=\rho_{i,i}$ with indices $i$ running as $b,1=a_1,2=a_2,3=a_3,\alpha,\beta$.
Similarly, the PMEs for the dipole-dipole coupled case, whose processes are shown in Fig.~\ref{states2} (b), 
are given by
\begin{equation}
\begin{split}
\label{e2}
\dot{p}_{1}
&= \gamma_{1h}[n_{1h}p_{b} -(1+n_{1h})p_{1}]
  +\gamma_{12}[n_{12}p_{2} -(1+n_{12})p_{1}]
  +\gamma_{13}[n_{13}p_{3} -(1+n_{13})p_{1}]
  +\gamma_{1c}[n_{1c}p_{\alpha} -(1+n_{1c})p_{1}],\\[5pt]
\dot{p}_{2}
&=\gamma_{12}[(1+n_{12}) p_{1} -n_{12}p_{2}]
 +\gamma_{23}[n_{23}p_{3} -(1+n_{23}) p_{2}],\\[5pt]
\dot{p}_{3}
&= \gamma_{3h}[n_{3h}p_{b} -(1+n_{3h}) p_{3}]
  +\gamma_{23}[(1+n_{23}) p_{2} -n_{23}p_{3}]
  +\gamma_{13}[(1+n_{13}) p_{1} -n_{13}p_{3}]
  +\gamma_{3c}[n_{3c}p_{\alpha}- (1+n_{3c}) p_{3}],\\[5pt]
\dot{p}_{\alpha}
&=\gamma_{1c}[(1+n_{1c}) p_{1} -n_{1c} p_{\alpha}]
 +\gamma_{3c}[(1+n_{3c}) p_{3} -n_{3c} p_{\alpha}]
 -\Gamma(1+\chi)p_{\alpha},\\[5pt]
\dot{p}_{\beta} 
&=\Gamma p_{\alpha} +\Gamma_c[N_c p_b - (1+N_c) p_{\beta}],
\end{split}
\end{equation}
\end{widetext}
where index $i$ of $p_i$ runs as $b,1=x_1,2=x_2,3=x_3,\alpha,\beta$. In Eqs.~{(\ref{e1})} and 
(\ref{e2}), the equation of motion for $p_b=\rho_{bb}$ is determined by the conservation of 
the probability, $\sum_ip_i =\sum_i\rho_{ii} = 1$.  In Eqs. (\ref{e1}) and (\ref{e2}), $n_h$ 
and $n_{1h}$ ($n_{3h}$) stand for the average numbers of photons with frequencies matching 
the transition energies from the ground state $\ket{b}$ to the excited states $\ket{a_i}$ 
and $\ket{x_1}$ ($\ket{x_3}$), respectively. $n_c$ and $n_{1c}$ ($n_{3c}$) 
are the thermal occupation numbers of ambient phonons at room temperature, $T_a = 300\text{\;K}$, 
with energies $E-E_\alpha$ in Eq.~{(\ref{e1})} and $E_{x_1}-E_\alpha$ ($E_{x_3}-E_\alpha$) 
in Eq. (\ref{e2}). $n_{12}$, $n_{13}$, and $n_{23}$ represent the corresponding thermal 
occupations at $T_a$ with energies $E_{x_1}-E_{x_2}$, $E_{x_1}-E_{x_3}$, and $E_{x_2}-E_{x_3}$, 
respectively. $N_c$ is the corresponding thermal occupation at $T_a$ with 
the energy $E_{\beta}-E_b$. The rates in Eqs. (\ref{e1}) and (\ref{e2}) obey local detailed 
balance and correctly lead to a Boltzmann distribution for the level population 
if the thermal averages for the photon and phonon reservoirs are set to a common temperature, such as room 
temperature. We consider the initial condition to be a fully occupied ground state, i.e., $\rho_{bb}(t=0) = 1$.

\subsection{Results}
\begin{figure}[]
\includegraphics[width=0.45\textwidth]{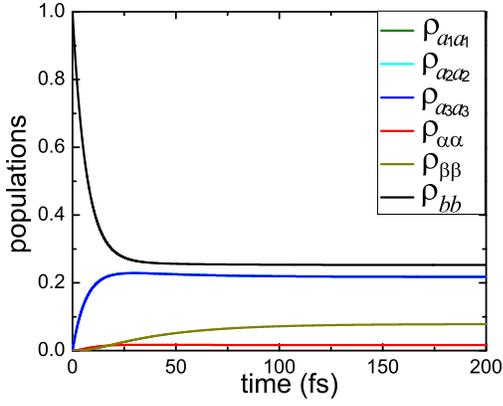}
\includegraphics[width=0.45\textwidth]{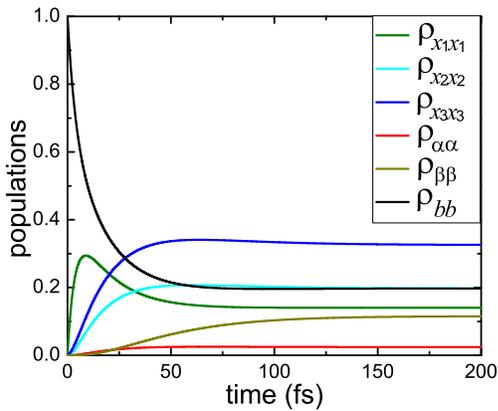}
\caption{(a) The time-evolutions of the populations $p_i$ of the levels 
from the numerical solution of the Pauli master equation (Eq. \ref{e1}) for uncoupled donors.
(b) Numerical solutions of the Pauli master equation (Eq. \ref{e2}) for coupled donors.}
\label{pop_dynamics}
\end{figure}

To calculate the population of each state, we use the following parameters: the energy levels 
are $E-E_b = 1.8\text{\ eV}$, $E-E_\alpha = E_\beta-E_b = 0.2\text{\ eV}$, 
$J_{12}=J_{23}=J=0.015\text{\ eV}$; the transfer rates are $\gamma_h = 0.62 \times 10^{-6}\text{\ eV}$, 
$\gamma_c = 6\text{\ meV}$, $\Gamma = 0.12\text{\ eV}$, $\Gamma_c = 0.025\text{\ eV}
$~\cite{chin,Scully2010,Dorfman2013}. 
We assume that the superposition states are stable under the steady-state operation, so that 
$\gamma_{13},\gamma_{12},\gamma_{23}$ have to satisfy the relationship: 
$\gamma_{13}=2\gamma_{12}=2\gamma_{23}\le2\sqrt{2}J$~\cite{rate}. 
Here, we choose as a limiting condition: $\gamma_{13}=2\gamma_{12}=2\gamma_{23}=2\sqrt{2}J$. 
We also employed this as a limiting condition for Creator's model, to create an appropriate 
comparison with our model. Figs.~\ref{pop_dynamics} (a) and (b) show the populations of each state in 
the absence and presence of coupling.  Due to the dipolar interaction among donors, the populations 
of the donors' ground state, $\ket{b}$, is significantly decreased while the populations of 
the acceptors' states, $\ket{\alpha}$ and $\ket{\beta}$, are notably increased in the presence 
of coherence when the system reaches the steady-state operation. These changes are responsible 
for the enhanced photocurrents. 

\begin{figure}[]
\includegraphics[width=0.45\textwidth]{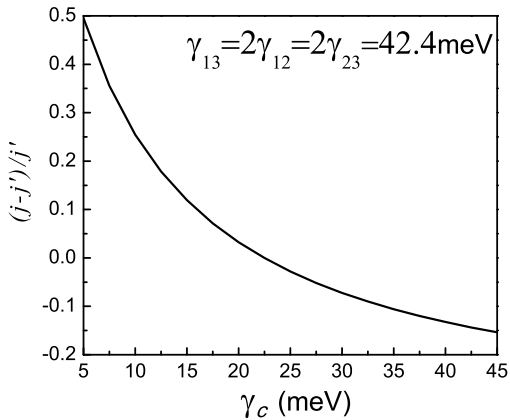}
\caption{Relative current enhancement $(j-j^\prime)/j^\prime$ as a function of the transition rate 
$\gamma_c$ between the donors and acceptor. Using the upper limit condition for $\gamma_{12}$, 
$\gamma_{23}$, and $\gamma_{13}$, we can get a current enhancement as high as 49.5\%. 
On the other hand when $\gamma_c = \gamma_{12} = \gamma_{23}$, there is no current enhancement.}
\label{eff}
\end{figure}

\begin{figure}[htpb]
\includegraphics[width=0.45\textwidth]{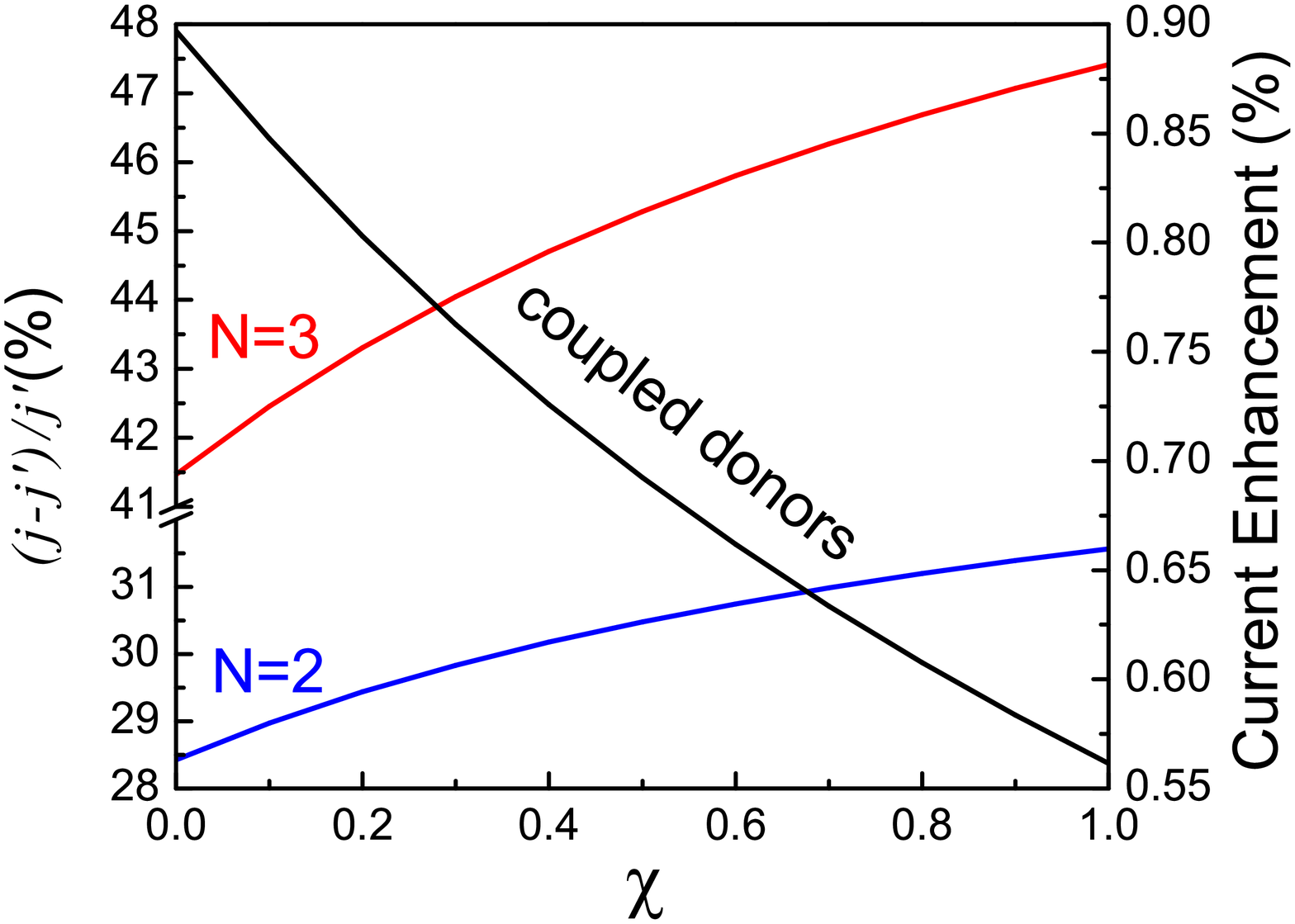}
\caption{Relative current enhancement $(j-j^\prime)/j^\prime$ as a function of the recombination 
rate $\chi$ using $\gamma_c = 6\text{\ meV}$. $j$ and $j^\prime$ are the electric current in 
the excitonically coupled and uncoupled cases, respectively, when the system reaches steady-state 
operation. The red line represents the current enhancement for the system with three donors; 
the blue line represents the current enhancement for the system with two donors proposed by 
Creator {\it et al.}; the black line represents the current enhancement of our model comparing 
to Creator's model in the presence of dipolar coupling.}
\label{chi}
\end{figure}

\begin{figure}[htbp]
\includegraphics[width=0.45\textwidth]{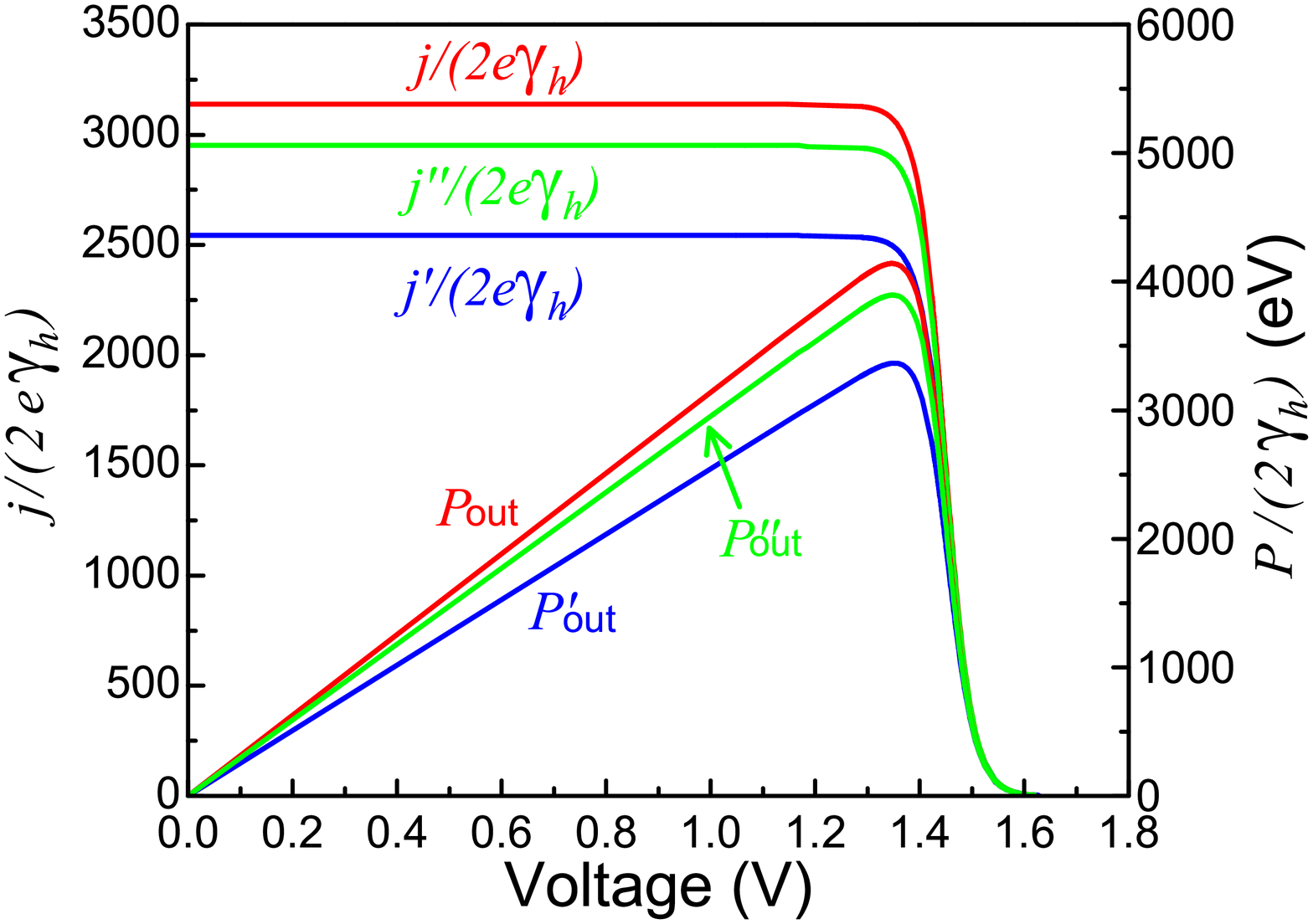}
\caption{Current and power generated as a function of the induced cell voltage, $V$, at room temperature. 
The blue lines represent the dimensionless current $(j^\prime/(2e\gamma_h))$ and power generated ($P^\prime_{out}$) 
of the system with three uncoupled dipoles ($J_{12}=J_{23}=J_{13}=0$); the red lines represent 
the dimensionless current $(j/(2e\gamma_h))$ and power generated ($P_{out}$) of the system with 
three coupled dipoles ($J_{12}=J_{23}\ne0,J_{13}=0$); the green lines represent the dimensionless 
current ($j^{\prime\prime}/(2e\gamma_h)$) and power generated ($P^{\prime\prime}_{out}$) of the system 
with two coupled dipoles, which is proposed in Ref.~\onlinecite{chin}.}
\label{power}
\end{figure}

Taking a modest recombination rate $\Gamma_{a\to b}=\chi\Gamma$ with $\chi = 20\%$,  
Fig.~\ref{eff} shows the current enhancement as a function of the transition rate, $\gamma_c$, 
using the other parameters listed before. Under the upper limit condition, when 
$\gamma_c = \gamma_{12} = \gamma_{23}$, there is no current enhancement. This means that the charge 
transfer via the channels $\ket{x_1} \to \ket{\alpha}$, $\ket{x_1} \to \ket{x_3} \to\ket{\alpha}$, 
and $\ket{x_1} \to \ket{x_2} \to \ket{x_3} \to \ket{\alpha}$ are as fast as the combined transfer 
through the independent channels $\ket{a_1} \to \ket{\alpha}$, $\ket{a_2} \to \ket{\alpha}$, and 
$\ket{a_3} \to \ket{\alpha}$. However, when $\gamma_c < \gamma_{12} = \gamma_{23}$, the coherent 
coupling leads to substantial current enhancements when compared to the configuration without coupling. 
Fig.~\ref{eff} also shows that the current enhancement may reach as high as 49.5\%, comparing this 
with 35\% in Creator's model. This can be explained by two factors: (i) the optical transition
rate between the ground and the bright state is enhanced from 2 times to  2.9 times. (ii)
the electron transition from the almost-dark (dark state in Ref.~\cite{chin}) 
to acceptor is increased from 2 times to 2.9 times. So a simple calculation shows the enhancement of 
PV model, $49.5\% \approx \frac{2.9}{2}\times 35\%$.

 
We have also explored the effect of the recombination rate, $\Gamma_{a\to b}=\chi\Gamma$, on the current 
enhancement. In Fig.~\ref{chi}, we show not only the current enhancement for the system comprised of three 
donors, (which is proposed here,) but also the current enhancement for the system with two donors, 
(that proposed in Ref.~\onlinecite{chin},) under similar electron transfer rate conditions. The results 
show that although the overall current is lower for faster recombination, the relative enhancement of 
the photocurrent is actually slightly larger for strong recombination. This behavior is analogous to that 
in the system with two donors~\cite{chin}. From Fig.~\ref{chi}, we also notice that the current enhancement 
in our three-donor system is much larger than that in Creatore's two-donor system at any value for 
the recombination rates. However, the current enhancement, on the order of $10^{-3}$, from 
the two-coupled-donor model to the three-coupled-donor model is very small.

Within this scheme, we assume there to be a {\it load} connecting the acceptor levels $\alpha$ and $\beta$. 
According to Fermi-Dirac statistics, the voltage, $V$, across this load can be expressed as 
$eV = E_\alpha-E_\beta+k_{\rm B}T_a\log(\rho_{\alpha\alpha}/\rho_{\beta\beta})$, where $e$ is the fundamental 
charge of the electron~\cite{chin,Scully2010}. Thus, we can assess the performance of our proposed 
photocell in terms of its photovoltaic properties. The steady-state current-voltage $(j-V)$ characteristic 
and power generated are shown in Fig.~\ref{power}. The current and voltage are evaluated using the steady-state 
solutions of the PMEs; the calculations are performed at increasing rate $\Gamma$ as other parameters are fixed, 
from open circuit regime where $j\to0$ ($\Gamma\to0$) to the short circuit regime where $V \to 0$. 
The power, $P$, is evaluated by the formula $P=j\cdot V$. In Fig.~\ref{power}, the peak current enhancement 
is roughly 23.4\% in uncoupled three-donor system ($J_{12}=J_{23}=J_{13}=0$) relative to the coupled three-donor 
system ($J_{12}=J_{23}\ne0,J_{13}=0$). Following the definition contained in Ref.~\cite{chin}, the peak delivered 
power enhancement is about 23.0\% in the uncoupled three-donor system ($J_{12}=J_{23}=J_{13}=0$) relative 
to the coupled three-donor system ($J_{12}=J_{23}\ne0,J_{13}=0$). We also show in Fig.~\ref{power} that 
the current-voltage characteristic and power generated for the system with two coherent donors, proposed 
in Ref.~\cite{chin}. From these results, we find that when compared to the two coherent donor system, 
the three coherent donor system has an enhancement of 6.3\% in both peak current and peak delivered power. 

\section{Conclusion}
The study of photosynthesis has inspired a new method by which we may harness quantum effects 
and coherent coupling amongst chromophores for the formation of coherent superposition 
to realize an artificial light-harvesting system at the molecular scale. In this paper, 
we propose a simple model to improve the performance of a theoretical photocell system. 
With suitably arranged electron donors, the photocurrents and power 
can be greatly enhanced through harnessing quantum effects.

The studied system is a photocell where the excitations are assumed resonant; for solar cells 
the excitation is done by solar radiation which has broad spectrum. However, the presented 
approach can be utilized in solar cells in different ways.
One approach is to extend the system into {\it N}-dipole (extended bands) and use solar 
radiation for excitation.
Another possibility is to host the dipole aggregates in solar cell materials close to the LUMO 
to suppress recombination and hence increase the collected photogenerated carriers~\cite{Saikin}.

Developing new concepts to harvest and utilize energy based on lessons learned
from nature like those in photosynthesis is of great current interest. Examining 
the current-voltage characteristic and power 
generated for the system with three coherent dipoles, we have found an
efficiency enhancement of about 6.3\% compared with two coherent dipoles. 
This encouraging  trend  suggest a promising novel 
design aspect of photosynthesis-mimicking photovoltaic devices.



\begin{thebibliography}{99}
\bibitem{note1} The Hamiltonian~(\ref{Hamil_exciton}) is identical to an XY spin Hamiltonian. One exciton state
  corresponds to the one spin flipped state.
\bibitem{note2} The single exciton eigenstates of Hamiltonian~(\ref{Hamil_exciton}) are given explicitly as 
      $\ket{\pm}=\frac{1}{\sqrt{2}}(\ket{a_1b_2} \pm \ket{b_1a_2})$. For simplicity,
      these are written as $\ket{\pm}=\frac{1}{\sqrt{2}}(\ket{a_1} \pm \ket{a_2})$.
\end{thebibliography}
\end{document}